\documentclass[preprint,1p,12pt]{elsarticle}

\usepackage{psfrag}

\usepackage{axodraw4j}
\usepackage{pstricks}
\usepackage{color}
\usepackage{hyperref} 
\usepackage{amsmath}
\usepackage{epsfig}
\usepackage{amsfonts}
\usepackage{amssymb}

\newcommand{\ie}{{\it i.e.}}
\newcommand{\eg}{{\it e.g.}}

\newcommand{\pythia}{{\sc Pythia}}
\newcommand{\pythiasix}{{\sc Pythia6}}
\newcommand{\herwigpp}{{\sc Herwig++}}
\newcommand{\herwig}{{\sc Herwig}}

\newcommand{\madloop}{{\sc MadLoop}}
\newcommand{\cuttools}{{\sc CutTools}}
\newcommand{\madfks}{{\sc MadFKS}}
\newcommand{\mcatnlo}{{\sc MC@NLO}}
\newcommand{\amcatnlo}{a{\sc MC@NLO}}

\begin{document}


\begin{flushright}
CERN-PH-TH/2011-091\\
CP3-11-17\\
ZU-TH 07/11
\end{flushright}
\vskip1cm

\title{Scalar and pseudoscalar Higgs production\\ 
in association with a top-antitop pair}

\author[1]{R.Frederix}
\author[2,3]{S.Frixione}
\author[3]{V.Hirschi}
\author[4]{F.Maltoni}
\author[2,5]{R.Pittau}
\author[3]{P.Torrielli}

\address[1]{ Institut f\"ur Theoretische Physik, Universit\"at Z\"urich,
Winterthurerstrasse 190, CH-8057 Z\"urich, Switzerland}
\address[2]{PH Department, TH Unit, CERN, CH-1211 Geneva 23, Switzerland}
\address[3]{ITPP, EPFL, CH-1015 Lausanne, Switzerland}
\address[4]{ Centre for Cosmology, Particle Physics and Phenomenology (CP3),
Chemin du Cyclotron 2, Universit\'e Catholique de Louvain, Belgium}
\address[5]{ Departamento de  F\'\i sica Te\'orica y del Cosmos y CAFPE,
  Universidad de Granada, Spain}

\begin{abstract}
We present the calculation of scalar and pseudoscalar Higgs
production in association with a top-antitop pair to the next-to-leading order
(NLO) accuracy in QCD, interfaced with parton showers according to 
the \mcatnlo\ formalism. 
We apply our results to the cases of light and very light Higgs boson
production at the LHC, giving results for total rates as well as for
sample differential distributions, relevant to the Higgs, to the top quarks,
and to their decay products. This work constitutes the first 
phenomenological application of \amcatnlo, a fully automated approach 
to complete event generation at NLO in QCD.
\end{abstract}

\maketitle

\newpage
\section{Introduction}

Establishing evidence for the Higgs boson(s), \ie, the scalar remnant(s) of
the Englert-Brout-Higgs mechanism~\cite{englert:1964et,higgs:1964ia,higgs:1964pj} in the standard
model and in extensions thereof, is among the most challenging goals of the
LHC experiments.  A coordinated theoretical/experimental effort in the last
years has led to a number of remarkable achievements in the accuracy and
usefulness of the available theoretical predictions, and in the role
these play in current analysis techniques~\cite{Dittmaier:2011ti}.

Depending on mass and couplings, Higgs bosons are produced and eventually decay
in a plethora of different ways, leading to a wide range of signatures. In
most cases, signals are difficult to identify because of the presence of large
backgrounds, and reliable predictions are necessary firstly to design efficient
search strategies, and secondly to perform the corresponding analyses. 
A particularly challenging
scenario at the LHC is that of a standard-model light Higgs, $m_H \lesssim
130$ GeV. In this case, the dominant decay mode is into a $b \bar b$ pair,
which is however completely overwhelmed by the irreducible QCD background.
A possible solution is that of considering the Higgs in association
with other easier-to-tag particles. An interesting case is that
of a top-antitop pair, since the large Yukawa coupling $ttH$, and the
presence of top quarks, can be exploited to extract the signal from 
its QCD multi-jet backgrounds. 
Unfortunately, this production mechanism is also plagued by large backgrounds 
that involve a $t\bar t$ pair, and hampered by its rather small
rates, and thus turns out to be difficult to single out. Several
search strategies have been proposed, based on different decay modes: from
$b \bar b$ which leads to largest number of expected events, to the more rare
but potentially cleaner $\tau\tau$~\cite{Belyaev:2002ua},
$WW^{(*)}$~\cite{Maltoni:2002jr} and $\gamma\gamma$~\cite{Buttar:2006zd} final
states.  All of them are in fact very challenging, and dedicated efforts
need be made.  For example, recently it has been argued that in the
kinematical regions where the Higgs is at quite high transverse momentum the
$b\bar b$ pair would be merged into one ``fat'' jet, whose typical structure 
could help in discriminating it from QCD 
backgrounds~\cite{Butterworth:2008iy,Plehn:2009rk} (boosted Higgs scenario).

It is then clear that accurate and flexible simulations, for both signals and
backgrounds, can give a significant contribution to the success of any
given analysis. Predictions accurate to NLO in QCD and at the parton
level for $t\bar t H$ hadroproduction have been known for some 
time~\cite{Beenakker:2001rj,Beenakker:2002nc,Dawson:2002tg,
Dawson:2003zu,Dittmaier:2003ej,Wu:2005gm}, and recently confirmed by other 
groups~\cite{Binoth:2010ra,Hirschi:2011pa}.
As for the most relevant background processes to the Higgs 
decay mode into $b \bar b$,
NLO calculations for ${t \bar t b \bar b}$~\cite{Bredenstein:2009aj,Bevilacqua:2009zn,Bredenstein:2010rs} and
${t \bar t j j}$~\cite{Bevilacqua:2010ve} are available in the literature. 
In this work, we extend the results for the signal to computing the associated
production $t\bar t A$ of a pseudo-scalar Higgs boson. All aspects
of the calculations we present here are fully automated. One-loop
contributions have been evaluated with \madloop~\cite{Hirschi:2011pa},
that uses the OPP integrand reduction method~\cite{Ossola:2006us} as 
implemented in \cuttools~\cite{Ossola:2007ax}.    
The other matrix-element
contributions to the cross sections, their phase-space subtractions 
according to the FKS formalism~\cite{Frixione:1995ms}, their combinations 
with the one-loop results, and their integration are performed by 
\madfks~\cite{Frederix:2009yq}. The validation of \madloop\
and \madfks\ in the context of hadronic collisions has been
presented in Ref.~\cite{Hirschi:2011pa}. For the sake of the present
work, we have also performed a dedicated comparison with the results of
Ref.~\cite{Dittmaier:2011ti} for the total $t\bar t H$ cross section, and 
found agreement at the permille level for several Higgs masses. 

We have also matched our NLO results with parton showers using the
\mcatnlo\ method~\cite{Frixione:2002ik}. This matching procedure has
also been completely automated, and this work represents the first 
application of the \mcatnlo\ technique to non-trivial processes
which were previously available only at fixed order and at the parton level -- in other
words, to processes not already matched to showers by means
of a dedicated, final-state-specific, software. What said above
also implies that our results are the first example of NLO computations 
matched to showers in which {\em all} ingredients of the calculation
are automated, and integrated in a unique software framework.

We remind the reader that the structure of the \mcatnlo\ short-distance
cross sections is the same as that of the underlying NLO computation,
except for a pair of extra contributions, called MC subtraction
terms. These terms have a factorised form, namely, they are essentially
equal to the Born matrix elements, times a kernel whose main property
is that of being process-independent. This is what renders it possible
the automation of the construction of the MC subtraction terms, and
ultimately the implementation of the \mcatnlo\ prescription.
We call \amcatnlo\ the code that automates the \mcatnlo\ matching,
and we defer its detailed presentation to a forthcoming
paper~\cite{amcatnlo:2011}.
\amcatnlo\ uses \madfks\ for phase-space generation and
for the computation of the pure-NLO short
distance cross section of non-virtual origin, and on top of that
it computes the MC subtraction terms. One-loop contributions may be taken
from any program which evaluates virtual corrections and is compatible
with the Binoth-Les Houches format~\cite{Binoth:2010xt};
as was said before, we use \madloop\ for the predictions given in this work.
The resulting \mcatnlo\ partonic cross sections are integrated 
and unweighted by MINT~\cite{Nason:2007vt}, or by 
BASES/SPRING~\cite{Kawabata:1995th}\footnote{These integrators have been 
modified by us, in order to
give them the possibility of dealing with both positive- and
negative-weighted events.}.
\amcatnlo\ finally writes a Les Houches file with MC-readable hard events 
(which thus includes information on particles identities and their 
colour connections).

\section{Results at the LHC}
We present selected results for total cross sections and distributions 
relevant to $t\bar t H/t \bar t A$ production at the LHC 
in three scenarios:
\begin{itemize}
\item[I.] Scalar $H$, with $m_H=120$ GeV;
\item[II.] Pseudoscalar $A$, with $m_A=120$ GeV;
\item[III.] Pseudoscalar $A$, with $m_A=40$ GeV;
\end{itemize}
where the Yukawa coupling to the top is always assumed SM-like, 
$y_t/\sqrt2  = m_t/v $. 

The three scenarios above allow one to compare the effects due the
different parity of the Higgs couplings on total rates as well as
on differential distributions. In this respect, it is particularly
interesting to consider the situation in which the Higgs boson 
is light and pseudoscalar, as is predicted
in several beyond-the-standard-model theories (see 
\eg~Refs.~\cite{Dermisek:2008uu,deVisscher:2009zb,Ellwanger:2009dp}).  
The main purpose of this section is that of studying the impact
of QCD NLO corrections at both the parton level and after shower and
hadronisation. For the numerical analysis we choose $\mu_F = \mu_R =
\left({m_T^t m_T^{\bar t} m_T^{H/A}}\right)^{\frac13}$, where $m_T =\sqrt{m^2 +
p_T^2}$ and $m_t^{pole}=m_t^{\overline{MS}}=172.5$ GeV. We have used LO and
NLO MSTW2008 parton distribution functions for the corresponding cross
sections. The parton shower in \amcatnlo\ has been performed with Fortran 
\herwig~\cite{Marchesini:1991ch,Corcella:2000bw,Corcella:2002jc}, 
version 6.520
 \footnote{We remind the reader that the \mcatnlo\ formalism has
been employed to match NLO results with \herwigpp~\cite{Bahr:2008pv} and, to
a lesser extent, with \pythia~\cite{Sjostrand:2006za} (see 
Ref.~\cite{Frixione:2010ra} and Ref.~\cite{Torrielli:2010aw} respectively).
The automation of the matching to these event generators
is currently under way.}.

\begin{figure}[h]
  \centering
   \includegraphics[width=0.8\textwidth]{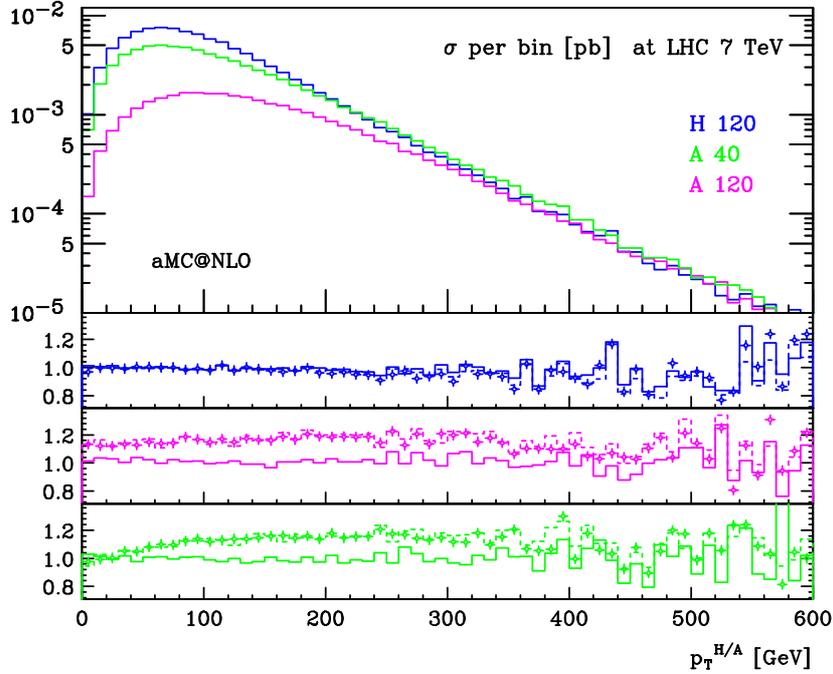}
    \caption{Higgs transverse momentum distributions in $t\bar t H/t \bar t
    A$ events at the LHC ($\sqrt{s}$=7 TeV), with \amcatnlo\ in the three
    scenarios described in the text: Scalar (blue) and pseudoscalar (magenta)
    Higgs with $m_{H/A}=120$ GeV and pseudoscalar (green) with $m_A=40$ GeV. In
    the lower panels, the ratios of \amcatnlo\ over LO (dashed), NLO (solid),
    and aMC@LO (crosses) are shown for each scenario.}
  \label{fig:one}
\end{figure}

\begin{figure}[h]
  \centering
   \includegraphics[width=0.8\textwidth]{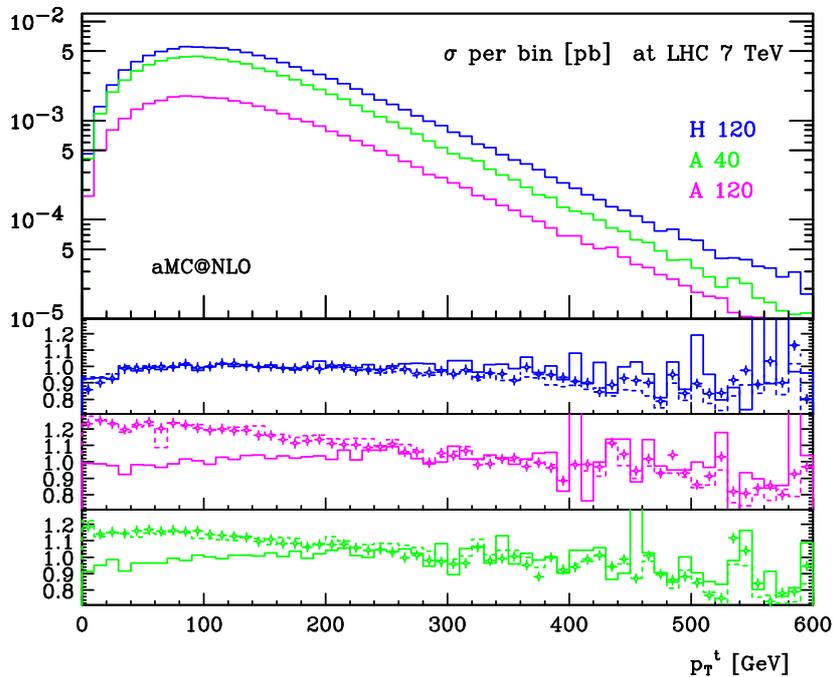}
    \caption{Same as in Fig.~\ref{fig:one}, for the $p_T$ of the top quark. }
  \label{fig:two}
\end{figure}

\begin{figure}[h]
  \centering
   \includegraphics[width=0.8\textwidth]{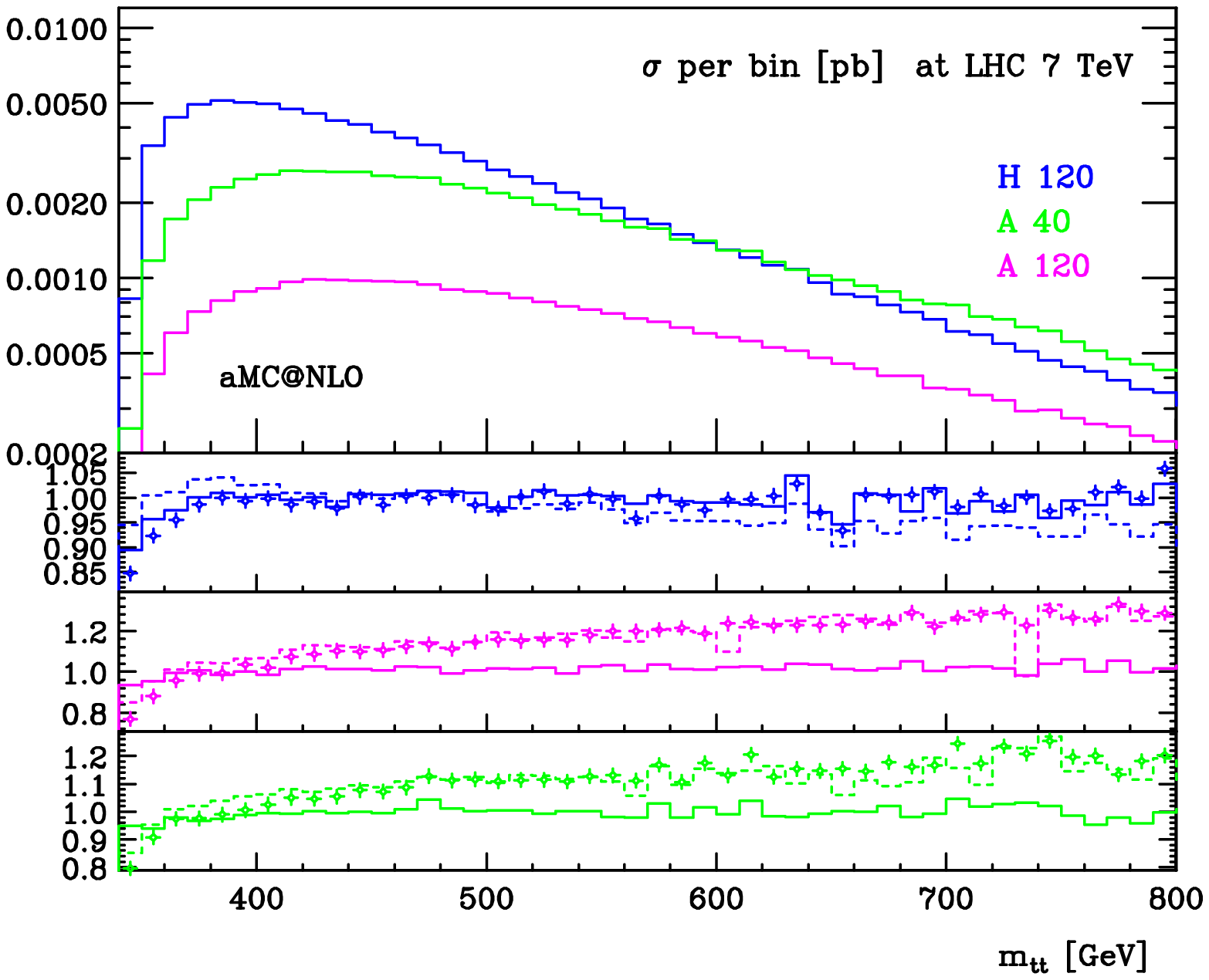}
    \caption{Same as in Fig.~\ref{fig:one}, for the invariant mass 
    of the top-antitop pair. }
  \label{fig:three}
\end{figure}

\begin{figure}[h]
  \centering
   \includegraphics[width=0.8\textwidth]{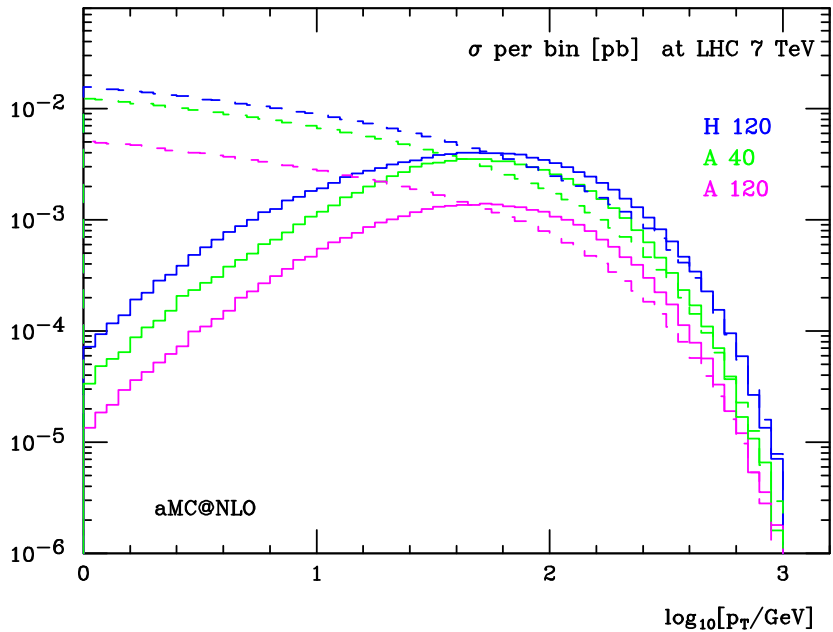}
    \caption{Transverse momentum of the $t\bar t H$ or $t\bar t A$
    system. The same colour patterns as in Fig.~\ref{fig:one}
    have been used. Solid histograms are \amcatnlo, dashed ones are NLO. }
  \label{fig:ptth}
\end{figure}

\begin{figure}[h]
  \centering
   \includegraphics[width=0.8\textwidth]{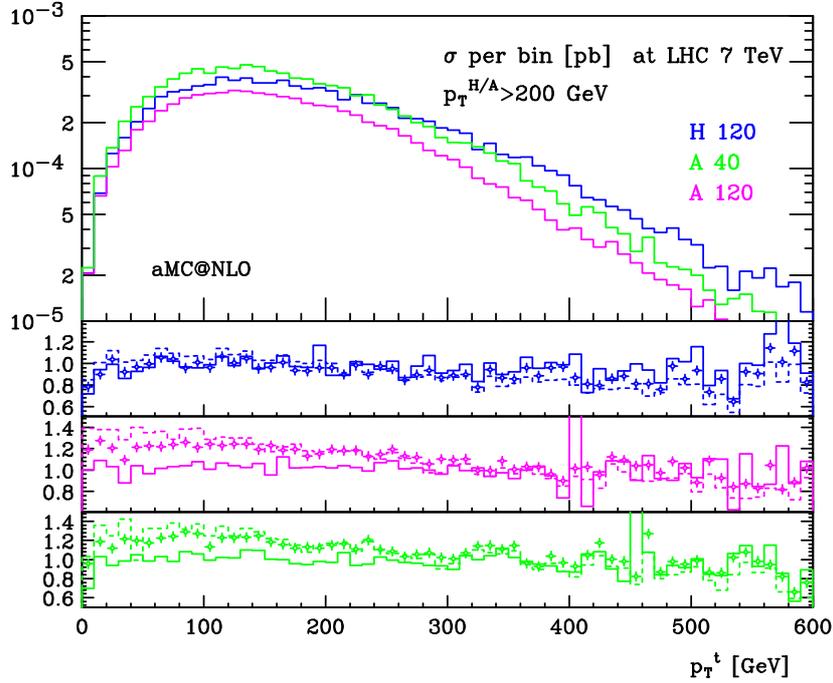}
    \caption{Same as in Fig.~\ref{fig:one}, for $p_T$ of top quark
    when $p_T^{H/A}>200$ GeV. }
  \label{fig:two-cut}
\end{figure}

\begin{figure}[h]
  \centering
   \includegraphics[width=0.8\textwidth]{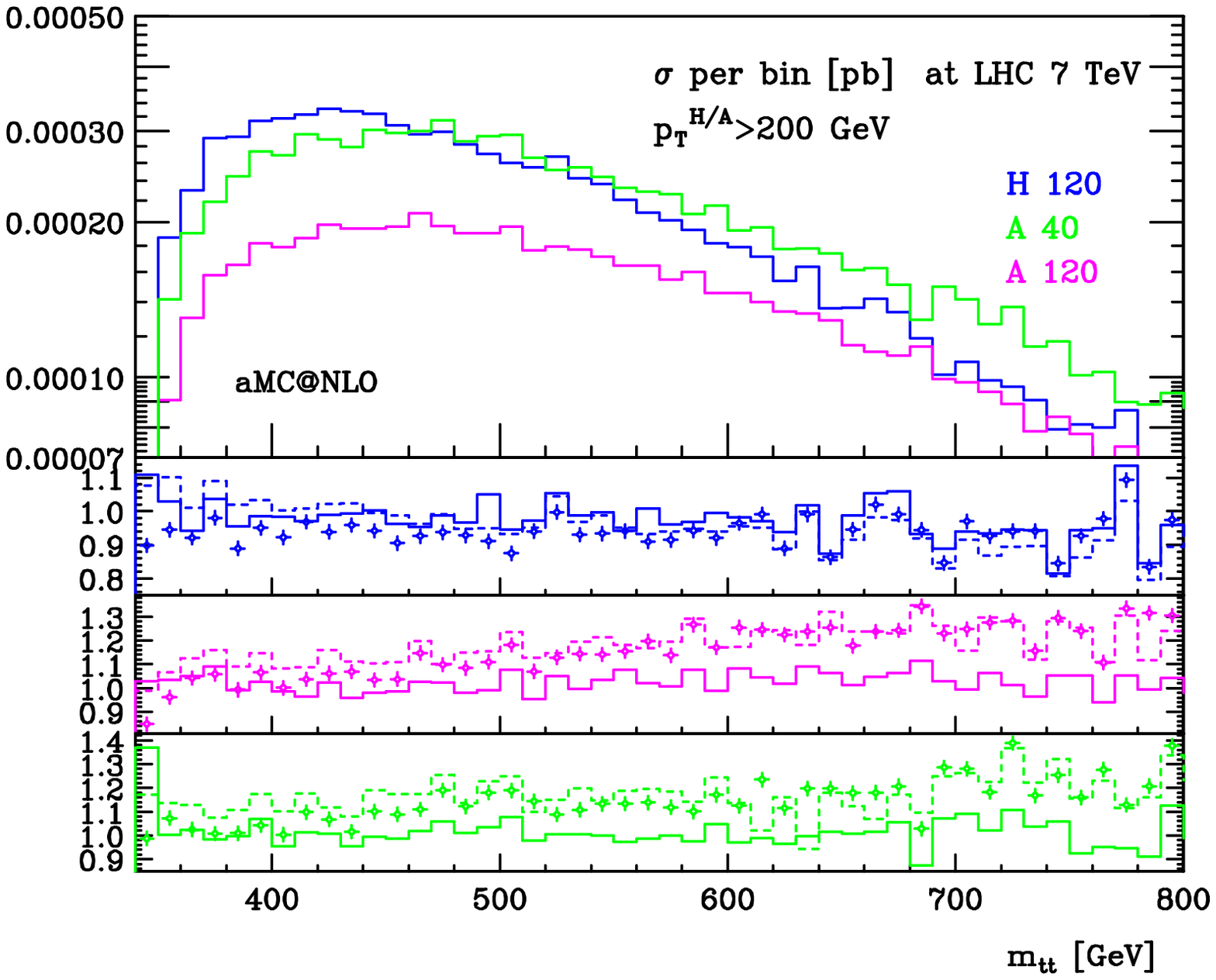}
    \caption{Same as Fig.~\ref{fig:two-cut}, for the invariant mass 
    of the top-antitop pair. }
  \label{fig:three-cut}
\end{figure}

\begin{figure}[h]
  \centering
   \includegraphics[width=0.6\textwidth]{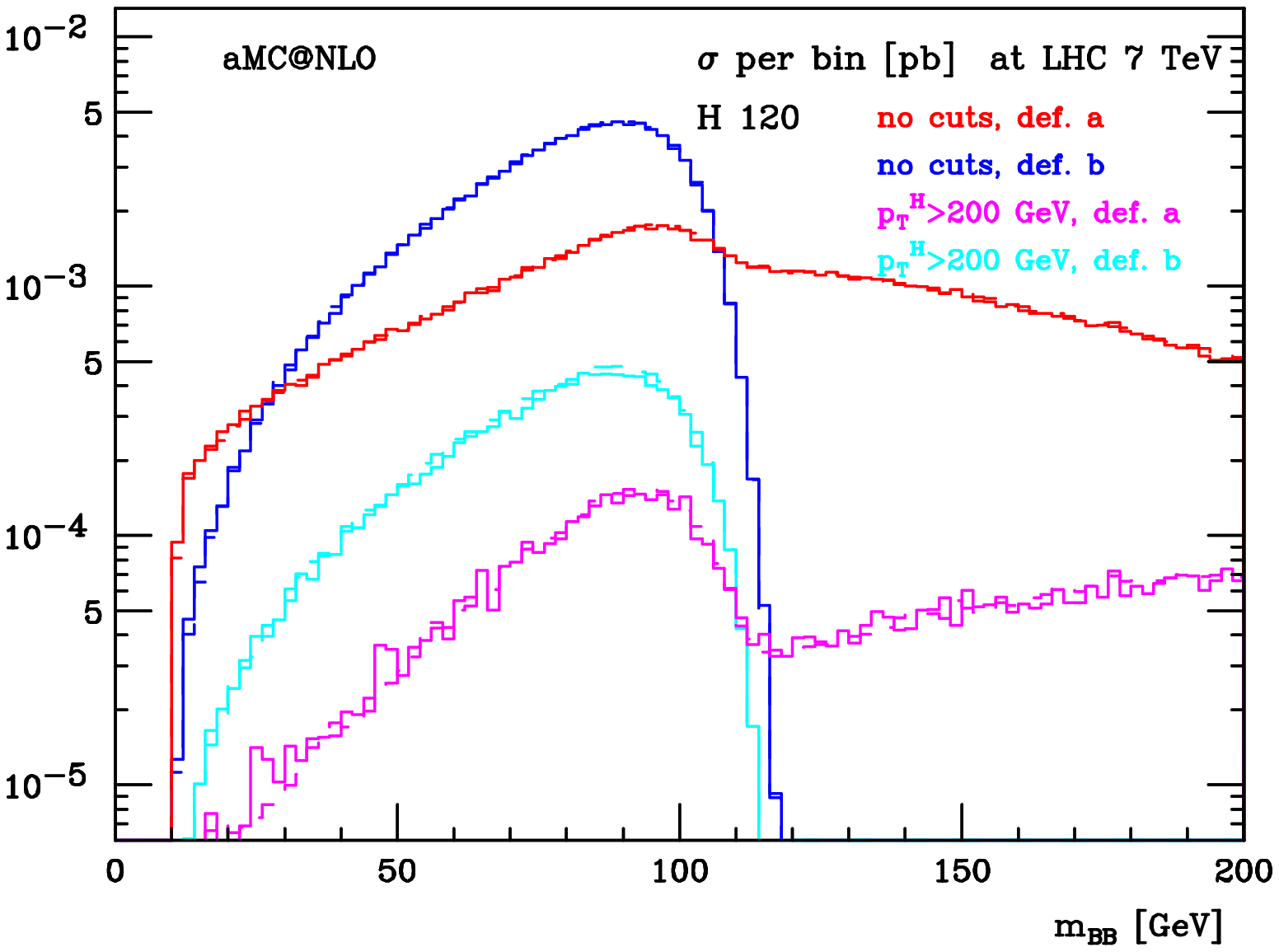}
   \includegraphics[width=0.6\textwidth]{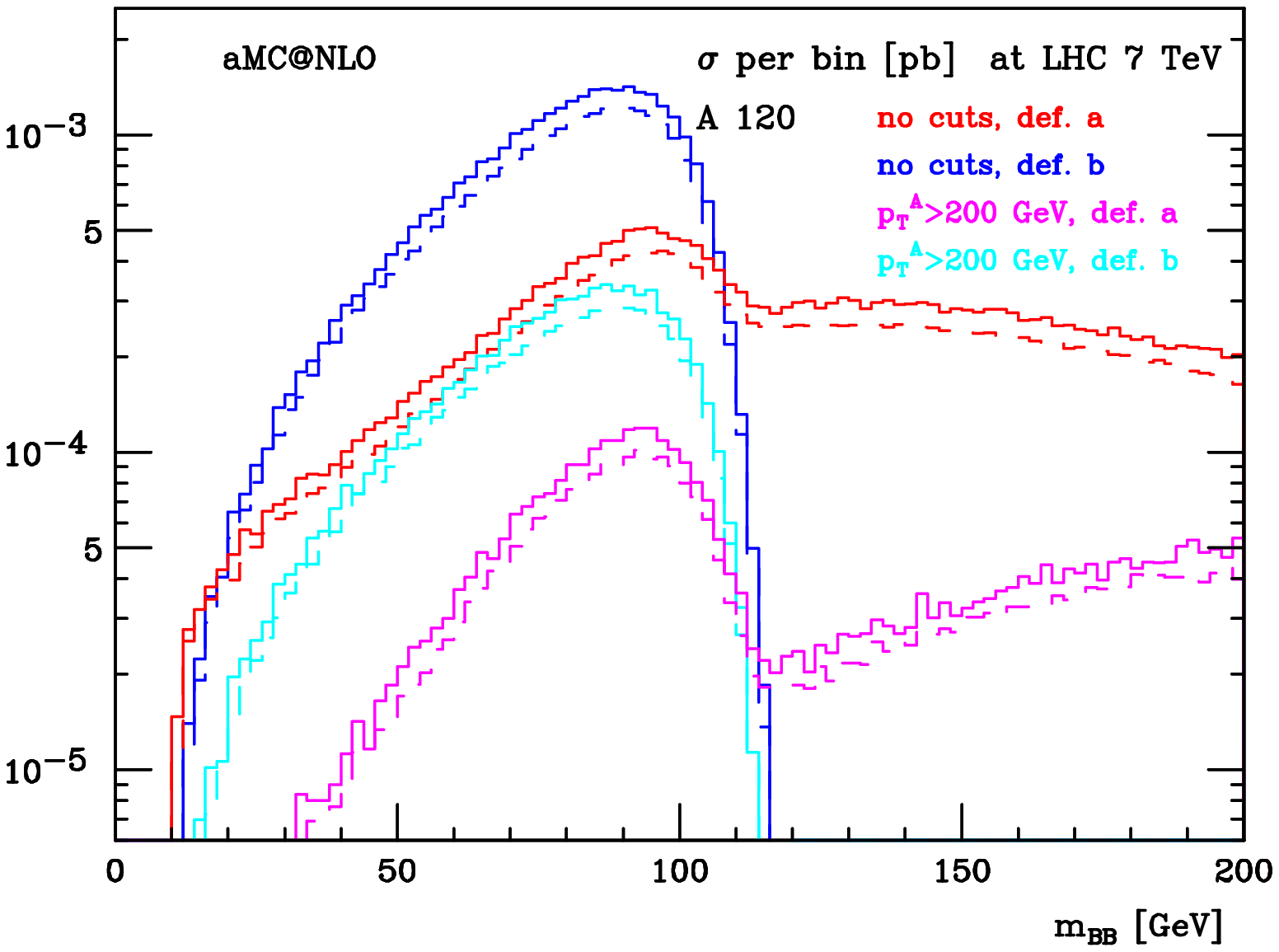}
   \includegraphics[width=0.6\textwidth]{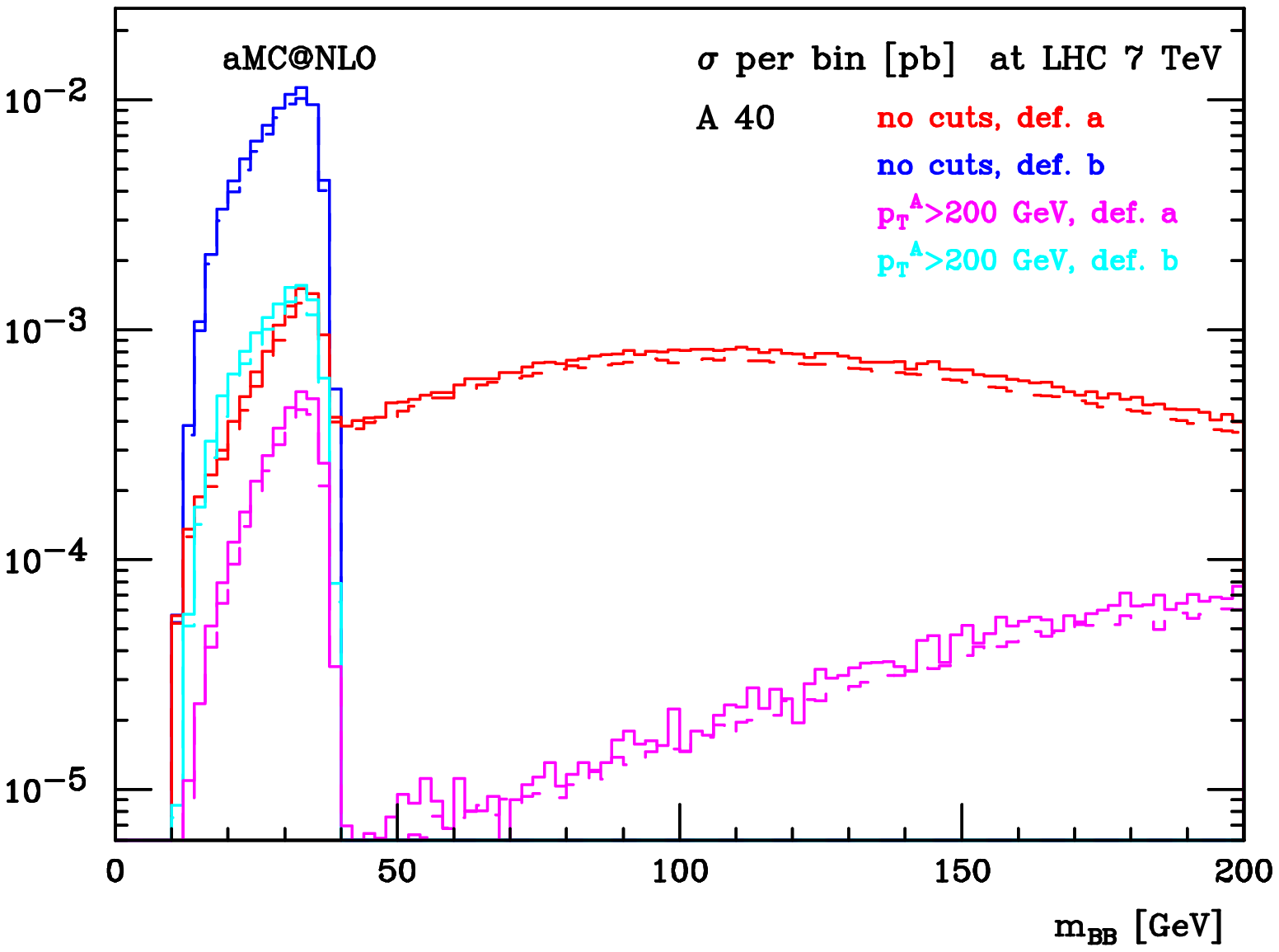}
    \caption{Invariant mass distributions of the $B$-hadron pairs
    defined as {\em a)} (red) and {\em b)} (blue) in the text.
    The results obtained by imposing $p_T^{H/A}>200$ GeV 
    (magenta and cyan, respectively) are also displayed.
    Solid histograms are \amcatnlo, dashed ones are aMC@LO. }
  \label{fig:mbb}
\end{figure}

\begin{figure}[h] 
  \centering
   \includegraphics[width=0.6\textwidth]{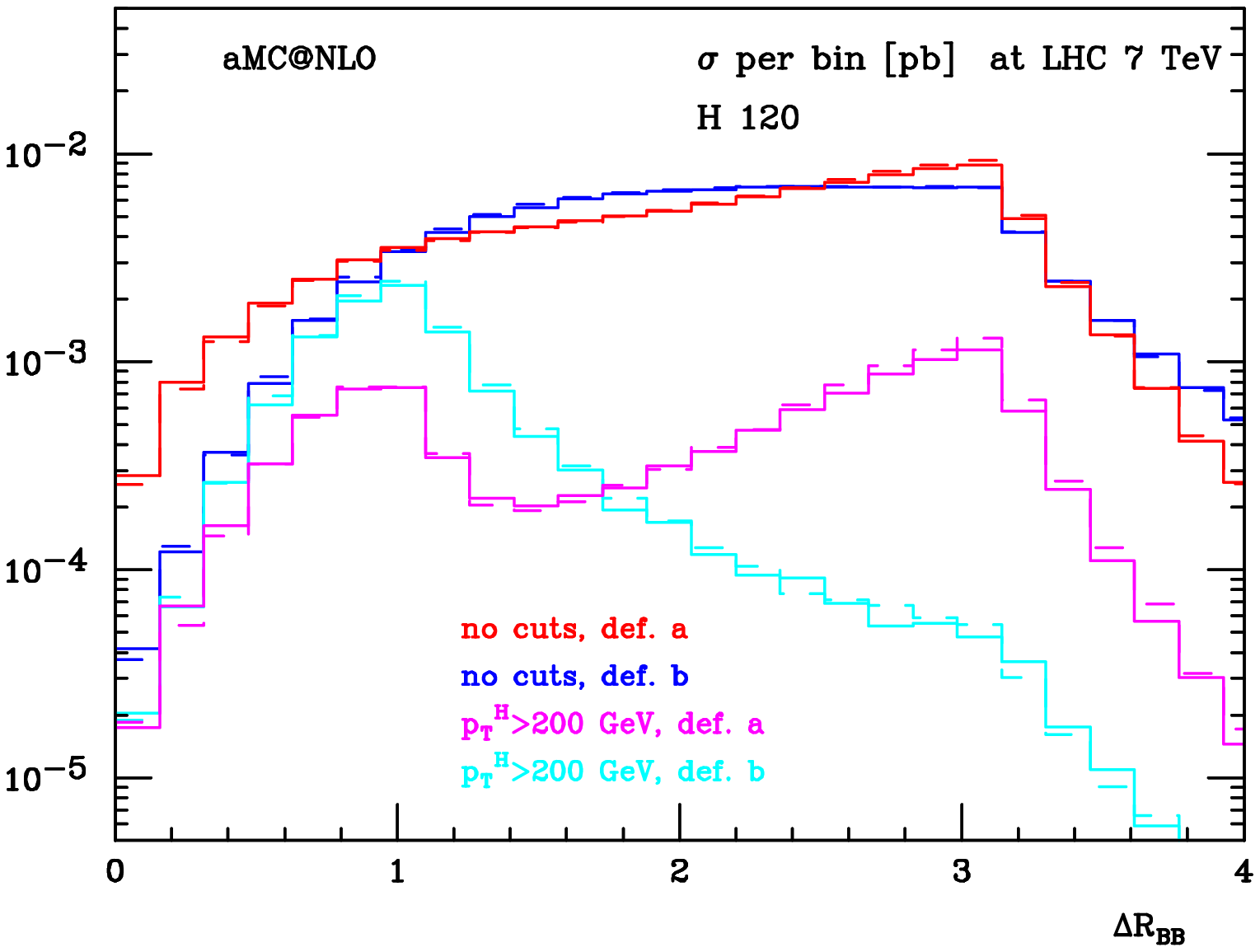}
   \includegraphics[width=0.6\textwidth]{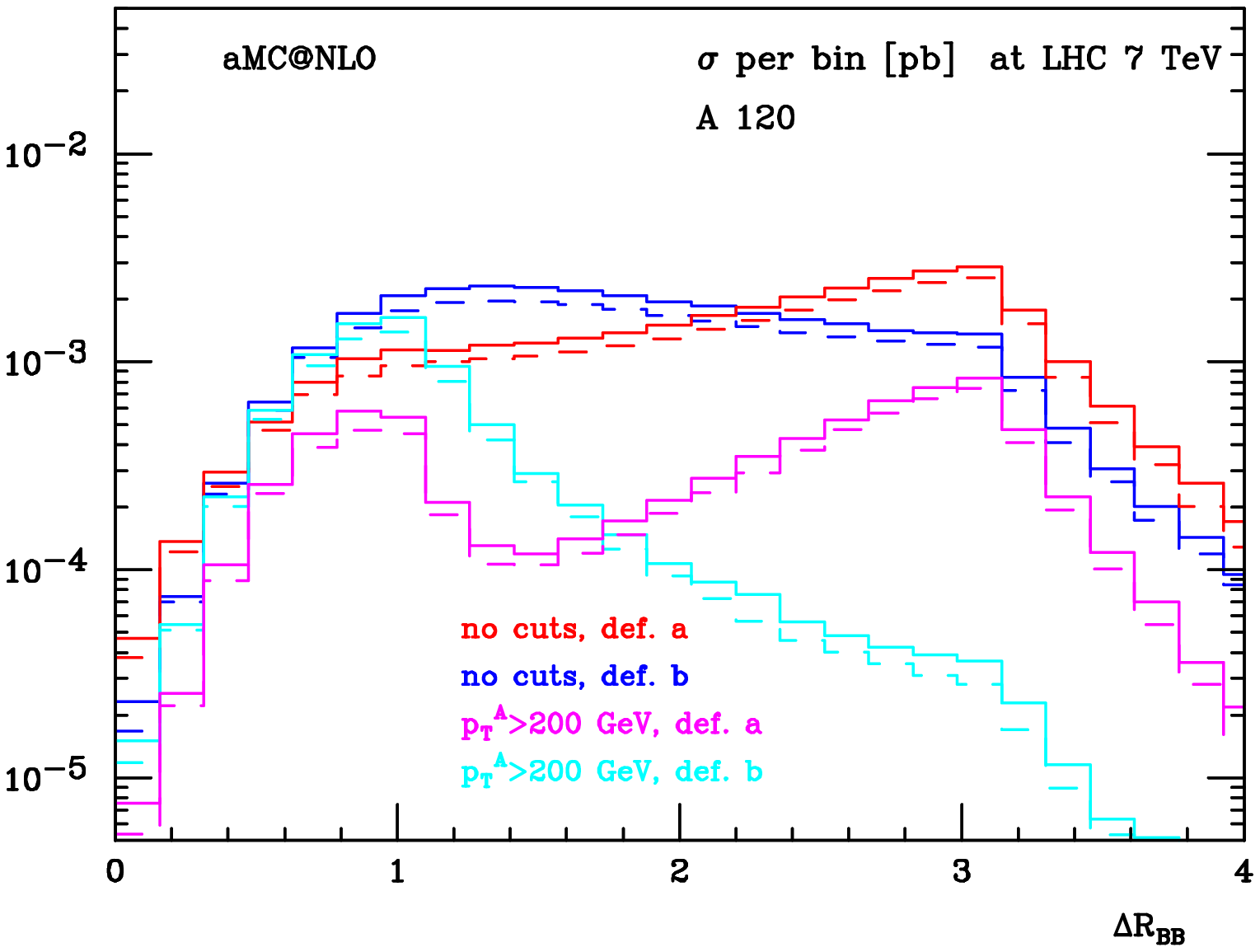}
   \includegraphics[width=0.6\textwidth]{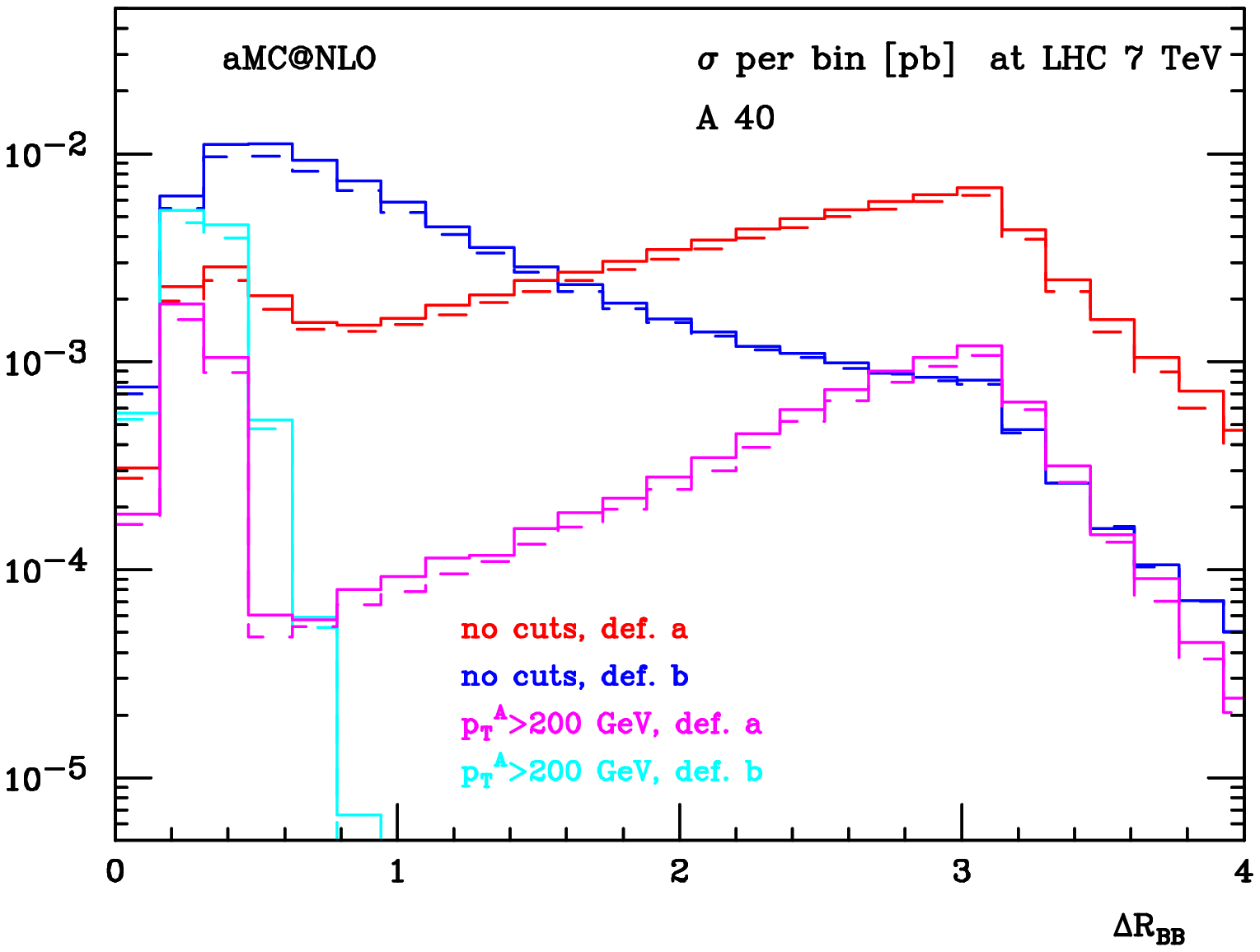}
    \caption{Same as in Fig.~\ref{fig:mbb}, for the $\Delta R_{BB}$ 
    correlation.
    }
  \label{fig:ddelRbb}
\end{figure}

The predicted production rates at the LHC running at $\sqrt{s}= 7\,{\rm
and}\,14$ TeV are given in Table~\ref{tab:xsec} where, for ease of reading,
we also show the fully inclusive $K$-factor. 
As far as differential distributions are concerned, 
we restrict ourselves to the 7~TeV LHC, and begin by studying
a few fully-inclusive ones (see Figs.~\ref{fig:one}-\ref{fig:ptth}).
We then consider a ``boosted'' case, \ie~apply a hard cut on the
transverse momentum of the Higgs (see Figs.~\ref{fig:two-cut} and
\ref{fig:three-cut}). Finally, in Figs.~\ref{fig:mbb} and \ref{fig:ddelRbb} 
we present our \amcatnlo\ predictions for correlations constructed with
final-state $B$ hadrons, which may or may not arise from the decays
of the Higgs and/or of the tops (see a discussion on this point later).

We first note a very interesting feature of Fig.~\ref{fig:one}: the $p_T$
distributions corresponding to the three different scenarios, while 
significantly different at small transverse momenta, 
become quite close to each other at higher values. This is expected from
the known pattern of the Higgs radiation off top quarks at high $p_T$ 
in both the scalar and the pseudoscalar cases~\cite{Dawson:1997im,Dittmaier:2000tc,Beenakker:2001rj}.
 This difference is not affected by NLO corrections, and could therefore be
exploited to identify the parity of the coupling at low $p_T$.  On the other
hand, the independence of the parity and masses of the $p_T$ distributions at
high values implies that the boosted analyses can equally well be used for 
pseudoscalar states.

In general, we find that differences between LO and aMC@LO\footnote{We
call aMC@LO the analogue of \amcatnlo, in which the short-distance cross 
sections are computed at the LO rather than at the NLO. Its results
are therefore equivalent to those one would obtain by using, \eg,
{\sc MadGraph/MadEvent}~\cite{Alwall:2007st} interfaced to showers.},
and between NLO and \amcatnlo, are quite small for observables 
involving single-inclusive distributions, see 
Figs.~\ref{fig:one}-\ref{fig:three}. The same remark applies to
the comparison between LO and NLO, and between aMC@LO and \amcatnlo.
However, if the cut $p^{H/A}_T >200$ GeV is 
imposed (boosted Higgs analysis), differences between LO and NLO 
(with or without showers) are 
more significant, and cannot be approximated by a constant $K$-factor.

As is obvious, the impact of the shower is clearly visible in the
three-particle $p_T(t \bar t H/t \bar t A)$ distribution of 
Fig.~\ref{fig:ptth}.
This observable is infrared-sensitive at the pure-NLO level for
$p_T\to 0$, where it diverges logarithmically. On the other hand,
the predictions obtained after interfacing with shower do display
the usual Sudakov suppression in the small-$p_T$ region.
At large transverse momenta the \amcatnlo\ and NLO predictions coincide 
in shape and absolute normalisation, as prescribed by the
\mcatnlo\ formalism.

\renewcommand{\arraystretch}{1.7}
\begin{table}[t]
\begin{center}
\begin{tabular}{|c|ccc|ccc|}
\hline
 & \multicolumn{6}{c|}{Cross section (fb)}\\

\cline{2-7}  
    Scenario  & \multicolumn{3}{c|}{7 TeV} &\multicolumn{3}{c|}{14 TeV} \\

 & LO &NLO &$K$-factor& LO & NLO & $K$-factor \\
\hline
I   & 104.5  &103.4 & 0.99  & 642  & 708 & 1.10\\
II  & 27.6   &31.9  & 1.16  & 244  & 289 & 1.18\\
III & 69.6   &77.3  & 1.11  & 516  & 599 & 1.16\\
\hline
\end{tabular}
\end{center}
\caption{Total cross sections for $t\bar t H$ and $t\bar t A$ production
at the LHC ($\sqrt{s}= 7,14$ TeV), to LO and NLO accuracy. The integration 
uncertainty is always well below 1\%.  Scale choices
and parameters are given in the text. }
\label{tab:xsec}
\end{table}

In our Monte Carlo simulations, we have included the $t\to e^+\nu b$, 
$\bar{t}\to e^-\bar \nu\bar{b}$, and $H\to b\bar{b}$ decays
at LO with their branching ratios set to one\footnote{We have neglected
production angular correlations~\cite{Frixione:2007zp}, as these are
expected to have a minor impact for the kind of processes and observables
we consider here. As usual when matching fixed order calculations to parton showers,
colour information is transferred in the large-$N_C$ limit.
The $b$-quark mass in the top and Higgs decay products has been set to the~\herwig~default, 4.95 GeV.}.
After showering, the $b$ quarks emerging from the
decays of the primary particles will result into $b$-flavoured
hadrons.
As prescribed by the MC@NLO formalism, the showering and hadronisation
steps are performed by the event generator the NLO computation is
matched to, \ie~\herwig~in this Letter. The parameters that control
hadron formation through cluster decays are set to their default
values in~\herwig~\cite{Corcella:2002jc}.
Additional $b$-flavoured hadrons may be produced as a consequence
of $g\to b\bar{b}$ branchings in the shower phase. For example, for 
scalar Higgs production at 7~TeV, about 2.7\% and 0.5\% of events have six 
and eight lowest-lying $B$ hadrons respectively. In our analysis, we have
searched the final state for all lowest-lying $B$ hadrons, and defined
two pairs out of them. {\em a)} The pair with the largest and 
next-to-largest transverse momenta; {\em b)} the pair with the largest 
and next-to-largest
transverse momenta among those $B$ hadrons whose parent parton was one
of the $b$ quarks emerging from the decay of the Higgs (there are about
0.2\% of events with four or six $B$ hadrons connected with the Higgs).
The definition of {\em b)} relies on MC truth (and in all cases we
assume 100\% tagging efficiency), but this is sufficient to study the
basic features of final-state $B$ hadrons.

In Figs.~\ref{fig:mbb} and \ref{fig:ddelRbb} we plot the  pair invariant mass
($m_{BB}$) and the $\eta-\varphi$ distance ($\Delta R_{BB}$) correlations 
between the $B$-hadron pairs
defined as explained above. The effects of the NLO corrections to $t\bar t H/t \bar t A$ are, 
in general, 
moderate. A cut of 200 GeV on the $p_T$ of the Higgs is seen to help
discriminate the $B$ hadrons arising from the Higgs from those coming
either from top decays, or from the shower.
The shapes of the distributions are similar between scenarios I and II
while, due to the lower Higgs mass, the $m_{BB}$ and $\Delta R_{BB}$ 
histograms peak at lower values in the case of a pseudoscalar $A$
with $m_A=40$ GeV.

\section{Conclusions}
Accurate and flexible predictions for Higgs physics will play an
important role in understanding the nature of the EWSB sector in the standard
model and beyond. In this Letter we have presented the results at NLO in QCD
for (scalar and pseudoscalar) Higgs production in association with a 
top-antitop quark pair, both with and without the matching to parton
showers.  Our approach is fully general and completely automated.  
A simple study performed on key observables involving the Higgs, the
top quarks, and their decay products shows that while changes in the 
overall rates can be up to almost $+$20\% (for the pseudoscalar states)
with respect to LO predictions, in general the shapes of distributions are 
mildly affected for a light SM Higgs. Significant changes, however, 
can be observed in the case of a light or very light pseudoscalar state.

The kernels of MC subtraction terms defined in the \mcatnlo\ formalism, 
although process-independent, do depend on the specific event generator
one adopts for the shower phase. In other words, each event generator 
requires a set of MC subtraction terms, which are computed analytically.
Results are now available for the cases of Fortran \herwig, \herwigpp, 
and \pythiasix; those relevant to the former program have been used to 
obtain the predictions presented here, while those relevant to the latter
two codes are presently being automated and tested against known
benchmarks.

We conclude by pointing out that work is in progress to make the use 
of \amcatnlo\ for $t\bar tH/t \bar tA$ production and for other
processes publicly available at {\tt http://amcatnlo.cern.ch}.

\section{Acknowledgments}
This research has been supported by the Swiss National Science Foundation
(NSF) under contract 200020-126691, by the IAP Program, BELSPO P6/11-P,
the IISN convention 4.4511.10 and by the Spanish Ministry of Education under contract PR2010-0285.
F.M. and R.P. thank the 
financial support of the MEC project FPA2008-02984 (FALCON).
S.F. is on leave of absence from INFN, Sezione di Genova, Italy.

\bibliographystyle{MyStyle}
\bibliography{biblio}

\end{document}